\documentclass[aps, 
amsmath, showpacs, preprintnumbers, twocolumn, amssymb, amsmath, superscriptaddress]{revtex4-1}
\usepackage{graphicx}
\usepackage{mathptmx, textcomp}
\usepackage[latin1]{inputenc}
\usepackage{tabularx}
\usepackage{units}
\usepackage{color}
\usepackage{bm}

\begin{document}
\title{Thermal broadening of the power spectra of laser-trapped particles in vacuum}
\author{M.\,Yoneda}
\affiliation{Department of Physics, Tokyo Institute of Technology, Ookayama 2-12-1, Meguro-ku, 152-8550 Tokyo}
\author{K.\,Aikawa}
\affiliation{Department of Physics, Tokyo Institute of Technology, Ookayama 2-12-1, Meguro-ku, 152-8550 Tokyo}

\date{\today}


\begin{abstract}
We show that at low pressures the spectral widths of the power spectra of laser-trapped particles are nearly independent from pressures and, due to the nonlinearities of the trap, reflect the thermal distribution of particles. In the experiments with nanoparticles trapped in an optical lattice, we identify two distinct features of the widths. First, the widths along an optical lattice are much broader than those in the other directions. Second, the spectral widths are narrower for larger nanoparticles. We develop a theory of thermal broadening and show that the spectral widths normalized by the frequencies of the center-of-mass motion directly reveal the ratio of the thermal energy to the trap depth. The presented model provides a good understanding of the observed features. Our model holds also for smaller particles such as atoms and molecules and can be readily extended to the general case with a single-beam optical trap.
\end{abstract}

\maketitle

\section{Introduction}\label{sec:intro}
Recent studies on optomechanical systems have shown remarkable progresses in controlling the motional state of nano- and micro-mechanical oscillators via light~\cite{aspelmeyer2014cavity}. These systems are promising platforms for diverse applications ranging from the investigation of macroscopic quantum phenomena~\cite{romero-isart2011large,pikovski2012probing,bassi2013models} and the sensitive detection of weak forces~\cite{stipe2001noncontact,moser2013ultrasensitive,gieseler2013thermal,ranjit2016zeptonewton} and small masses~\cite{yang2006zeptogram,chaste2012nanomechanical} to the studies on non-equilibrium physics at the nanoscale~\cite{millen2014nanoscale,dechant2015all-optical}. Among various optomechanical systems, optically levitated nanoparticles have a salient advantage that they are not in touch with any other objects and can potentially have an extremely high quality factor of $10^{12}$~\cite{li2011millikelvin,gieseler2012subkelvin,romero2010toward,chang2010cavity}. Great efforts have been made to cool the center-of-mass motion of nanoparticles and have demonstrated temperatures of the order of $\unit[1]{mK}$, close to its quantum ground state~\cite{li2011millikelvin,gieseler2012subkelvin,millen2015cavity,fonseca2016nonlinear,jain2016direct,vovrosh2017parametric}. 

In studies with optomechanical systems, a power spectral density (PSD) of the motion of the oscillator plays a central role in understanding and controlling its behavior. It has been well known that the profile of a PSD of levitated particles is given by a Lorentzian function, with spectral widths being the damping rate determined by the background pressure~\cite{epstein1924resistance,li2011millikelvin,gieseler2012subkelvin,vovrosh2017parametric}. Because the relation between the pressure and the damping rate depends on the mass of particles, the spectral widths extracted from PSDs have served as a crucial means to estimate the mass (and thus the radius) of trapped nanoparticles. 

In this article, we show that the spectral widths of the PSDs at pressures lower than tens of Pa are nearly constant and determined dominantly by the thermal broadening. Accordingly, the spectral profiles at low pressures significantly deviate from the Lorentzian function. The observed spectral widths at low pressures strongly depends on the direction of the motion as well as on the size of the nanoparticles. These behaviors are caused by the nonlinearities in the optical dipole force, which lower the frequency of the center-of-mass motion (trap frequency) under finite oscillation amplitudes. We identify two physical mechanisms for broadening: the anharmonicity and the cross-dimensional couplings. The former effect means a deviation of an optical potential from a harmonic potential, while the latter effect means that finite oscillation amplitudes in orthogonal directions effectively lowers the trap depth and hence the trap frequency. We derive a theoretical model accounting for both effects and show that the spectral width normalized by the trap frequency is directly connected to the ratio of the thermal energy to the trap depth. Our model explains the observed features well. With a slight modification on numerical coefficients representing nonlinearities, our model can be readily extended to a more general case with a single-beam optical trap.

The present study is closely related to the study on thermal nonlinearities~\cite{gieseler2013thermal}, where they investigated the shift of the trap frequency introduced by the nonlinearities of the optical force. In most part of their study, they applied feedback cooling to the orthogonal motions, which means canceling the cross-dimensional coupling, and focused on the one-dimensional problem of how the anharmonicity of the trap shifted the trap frequency. They did not explicitly explore the spectral widths. By contrast, the present study provides detailed investigations of the widths without applying feedback cooling and therefore deals with the three-dimensional problem. We also provide a new perspective on the relation between the width and the particle size.


\section{Experimental setup}\label{sec:exp}

\begin{figure}[t]
\includegraphics[width=0.95\columnwidth] {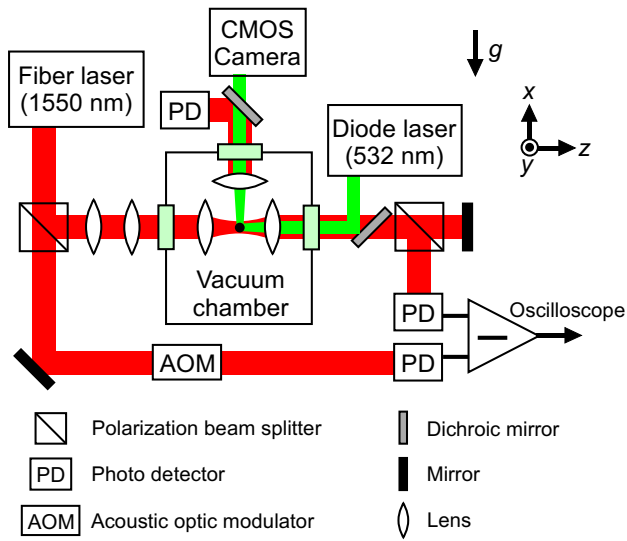}
\caption{(color online). Schematic of the experimental setup. A single-frequency fiber laser at $\unit[1550]{nm}$ is retro-reflected to form a one-dimensional optical lattice. A cylindrical lens pair, inserted in front of the chamber, is used for making an elliptical beam such that the trap frequencies in $x$ and $y$ directions are sufficiently separated. The three-dimensional center-of-mass motion of the nanoparticle is measured via a balanced photodetector. The nanoparticles are imaged on a CMOS camera from the side viewport. For imaging, we use a green diode laser at $\unit[532]{nm}$ overlapped with the trapping beam. The side viewport is also used for extracting the fluorescence from the nanoparticles, namely, the nanoparticles' Rayleigh scattering of the trapping beam. The acoustic optic modulator in front of the lower photodetector is used for precisely adjusting the offset of the balanced signal. The vacuum chamber is evacuated by a turbo molecular pump and a scroll pump that are not shown in the figure. }
\label{fig:setup}
\end{figure}

Our setup is based on previous experimental studies on feedback cooling the center-of-mass motion of nanoparticles optically levitated in vacuum~\cite{gieseler2012subkelvin,vovrosh2017parametric}. Here we briefly describe our setup. The schematic of our experimental setup is shown in Fig.~\ref{fig:setup}. A single-frequency fiber laser at a wavelength of $\unit[1550]{nm}$ with a power of about $\unit[300]{mW}$ is used for both trapping and detecting the motion of trapped nanoparticles. A high-numerical-aperture lens focuses the trapping beam to beam waists of about $\unit[1.7]{\mu m}$ and $\unit[2.0]{\mu m}$ in $x$ and $y$ directions, respectively. Another lens placed after the focusing lens is used for both collimating the trapping beam and collecting the scattered light from nanoparticles. The trapping beam is retro-reflected to form a standing-wave trap, namely, a one-dimensional optical lattice commonly used in cold atom experiments~\cite{bloch2005ultracold}. As a result, the trap frequency in the $z$ direction is much higher than in the other directions; the trap frequencies are about 36, 54, and $\unit[230]{kHz}\times 2\pi$ in the $x$, $y$, and $z$ directions, respectively.The three-dimensional motion of trapped nanoparticles is observed via a balanced photodetector, where the signal of the light without nanoparticles is subtracted from that of nanoparticles. For each measurement, we record the signal for $\unit[1]{s}$. The PSDs are obtained by performing a Fourier transform to the recorded data on a computer. The time duration for the Fourier transform is chosen to be $\unit[200]{ms}$, which is sufficiently longer than the Fourier limit of the observed narrowest spectrum (about $2\pi \times \unit[50]{Hz}$). From a single data set with a duration of $\unit[1]{s}$, we obtain five independent PSDs and extract the spectral widths from each. The standard deviation of the five values allows us to estimate errors in fitting.

\begin{figure}[t]
\includegraphics[width=0.5\columnwidth] {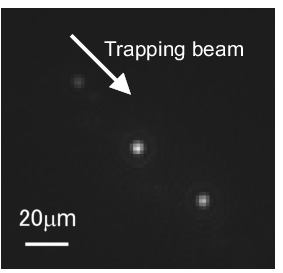}
\caption{Image of nanoparticles simultaneously trapped in an optical lattice. A standing wave trap is formed along the trapping beam. Nanoparticles are trapped at separate lattice sites. An imaging beam at $\unit[532]{nm}$ is overlapped with the trapping beam. For each experimental run, the number of trapped nanoparticles and their positions are different. In the presented image, three nanoparticles with different sizes are trapped.}
\label{fig:image}
\end{figure}

The Rayleigh-scattered light from trapped nanoparticles is collected also through the upper viewport. The scattering of an imaging beam at $\unit[532]{nm}$, overlapped with the trapping beam, is shone on a CMOS camera for acquiring the information on the number and the positions of trapped nanoparticles (Fig.~\ref{fig:image}). The magnitude of the scattering of the trapping beam is measured with an additional photodetector. This signal, which hereafter we call a fluorescence signal, provides a reliable means to estimate the relative size of trapped nanoparticles. In the present study, the observed fluorescence signals are ranged between 0.04 and $\unit[50]{\mu W}$. For loading of nanoparticles into the trap region, we introduce a mist of ethanol including silica nanoparticles with radii of about $\unit[60]{nm}$ into a vacuum chamber at atmospheric pressure. After we observe trapped nanoparticles on the camera, we evacuate the chamber to arbitrary pressures. Below $\unit[0.6]{Pa}$, nanoparticles escape from the trap. Hence, we investigate the spectra at pressures above $\unit[0.6]{Pa}$. 

The main feature of our setup lies in the optical lattice structure of the trap region. As a result, we are able to trap multiple nanoparticles at the same time as shown in Fig.~\ref{fig:image}. Trapping multiple nanoparticles in a standing-wave trap, formed by two counter-propagating beams emitted from two optical fibers, has been reported~\cite{monteiro2013dynamics}. In our setup, we are able to adjust the position of the trapped nanoparticles by controlling the position of the retro-reflecting mirror. Such a feature is necessary for the two following reasons. First, during the evacuation process, the position of trapped nanoparticles has to be adjusted to compensate a slight variation of the refractive index from that of air to that of vacuum, because such a variation in the refractive index shifts the standing wave by about $\unit[50]{\mu m}$ in our setup. Second, by pushing nanoparticles away from the focus of the beam, we are able to release them. In this way, we control number of remaining nanoparticles relevant to experiments. In the present study, we prepare a situation with only single nanoparticles trapped at the focus of the beam.

\section{Theoretical description of the thermal broadening}\label{sec:theory}

A previous study on thermal nonlinearities in the motion of trapped nanoparticles mainly focuses on a Duffing nonlinearity arising from the anharmonicity of the Gaussian distribution of a laser beam~\cite{gieseler2013thermal}. Although their theoretical framework starts with a general three dimensional problem, they simplify it to a one-dimensional problem because in their study the motions in the orthogonal two directions are suppressed by feedback cooling. On the contrary, we are interested in a situation where no feedback cooling is applied. Here we extend the formalism in ref.~\cite{gieseler2013thermal} such that both the anharmonicity and the cross-dimensional couplings are considered.

The purpose of this section is to find expressions of the thermal spectral widths on the basis of three dimensional equations of motion.  We first derive an expression of the optical dipole force taking into account nonlinearities to the first order. The spatial distribution of the optical potential created by a one-dimensional optical lattice beam is approximately given by~\cite{grimm2000optical}
\begin{equation}\label{eq:U}
U(x,y,z) \approx -U_0 \cos^2\left(\frac{2\pi z}{\lambda}\right){\rm exp}\left(-\frac{2x^2}{w_x^2}-\frac{2y^2}{w_y^2}\right)
\end{equation}
where $U_0$ is the depth of the potential, $\lambda$ is the wavelength of the beam, $w_x$ and $w_y$ are beam waists in $x$ and $y$ directions, respectively. Here, we assumed that the confinement in the $z$ direction dominantly arises from the standing wave and  ignored the intensity variation due to a beam divergence. The trap frequencies, without taking into account nonlinearities, are written as follows:

\begin{align}\label{eq:trapfreq}
\Omega_x = \sqrt{\dfrac{4U_0}{mw_x^2}},  
\Omega_y = \sqrt{\dfrac{4U_0}{mw_y^2}}, 
\Omega_z = \sqrt{\dfrac{8\pi^2U_0}{m\lambda^2}} 
\end{align}

The optical dipole force exerted by the trapping beam is obtained by differentiating eq.~(\ref{eq:U}) and can be approximated to the first order by
\begin{eqnarray}\label{eq:F}
{\bm F}\left(x,y,z\right) \approx U_0\left( 
\begin{array}{ccc}
-\dfrac{4x}{w_x^2} \left( 1-\dfrac{2x^2}{w_x^2}-\dfrac{2y^2}{w_y^2}-\dfrac{4\pi^2z^2}{\lambda^2} \right) \\
-\dfrac{4y}{w_y^2} \left( 1-\dfrac{2x^2}{w_x^2}-\dfrac{2y^2}{w_y^2}-\dfrac{4\pi^2z^2}{\lambda^2} \right) \\
-\dfrac{8\pi^2z}{\lambda^2} \left( 1-\dfrac{2x^2}{w_x^2}-\dfrac{2y^2}{w_y^2}-\dfrac{8\pi^2z^2}{3\lambda^2} \right) 
\end{array}
\right)
\end{eqnarray}
We define nonlinear coefficients $\xi_k^{(j)}$ such that the $j$-th component of the force (\ref{eq:F}) is expressed by them as follows: 
\begin{eqnarray}
F_j\left(x,y,z\right)  = -l_j \left( 1+ \sum^{}_{k} \xi_k^{(j)} x_k^2 \right) x_j
\end{eqnarray}
where $l_j$ is a spring constant in the $j$-th direction and $\{k,j\}\in\{x,y,z\}$. The nonlinear coefficients are then related to the beam waists and the wavelength as given in Table~\ref{tab:nlc}.

\begin{table}
\caption{\label{tab:nlc} Nonlinear coefficients expressed by the beam waists and the wavelength. }
\begin{ruledtabular}
\begin{tabular}{cccc}
$j$ & $\xi _j^{(x)}$ & $\xi _j^{(y)}$ & $\xi _j^{(z)}$ \\ \hline
$x$ & $-\dfrac{2}{w_x^2}$ & $-\dfrac{2}{w_x^2}$ & $-\dfrac{2}{w_x^2}$ \\
$y$ &  $-\dfrac{2}{w_y^2}$ & $-\dfrac{2}{w_y^2}$ & $-\dfrac{2}{w_y^2}$ \\
$z$ & $-\dfrac{4\pi^2}{\lambda^2}$ & $-\dfrac{4\pi^2}{\lambda^2}$ & $-\dfrac{8\pi^2}{3\lambda^2}$ \\
\end{tabular}
\end{ruledtabular}
\end{table}

Note that, for dealing with the case of a single-beam trap, only the nonlinear coefficients in the $z$ direction have to be modified. With modified coefficients, the following arguments on the spectral profiles and the spectral widths of PSDs hold also for a single-beam trap.

Using the nonlinear coefficients in Table~\ref{tab:nlc}, we can write down the three dimensional equations of motion as follows:
\begin{align}
\label{eq:motion}
\Ddot{q_j}&+\Gamma_0 \Dot{q_j} \notag \\ &+\Omega_j^2\left(1+\epsilon_j \cos \omega_j t + \sum_k\xi_k^{(j)} q_k^2 \right)q_j \approx 0
\end{align}
where $\epsilon_j$ and $\omega_j$ are the amplitude of parametric modulation and the modulation frequency of parametric driving in the $j$-th direction, respectively. The terms of parametric driving are necessary to find actual trap frequencies in the presence of thermal motions~\cite{gieseler2013thermal}, whereas terms relevant to feedback included in ref.~\cite{gieseler2013thermal} are omitted here. The damping coefficient $\Gamma_0$ is proportional to the background pressure $P$~\cite{beresnev1990motion}:

\begin{eqnarray}\label{eq:gamma}
\Gamma_0 = 0.619\dfrac{9\pi\phi d_m^2}{\sqrt{2}a\rho k_BT_{\rm air}}P
\end{eqnarray}
where $a$ is the radius of nanoparticles, $d_m$ is the diameter of air molecules, $\rho$ is the density of nanoparticles, $\phi$ is the viscosity of air, $k_B$ is the Boltzmann constant, and $T_{\rm air}$ is the temperature of the surrounding gas. 

Following the approach of ref.~\cite{gieseler2013thermal}, based on the secular perturbation theory, we introduce a dimensionless slow time scale $\tau$ given by
\begin{align}
\begin{split}
\tau=&\kappa_j\Omega_j t \\
\kappa_j\equiv&\dfrac{\Gamma_0}{\Omega_j}
\end{split}
\end{align}
where $\kappa_j$ corresponds to the inverse of the quality factor of the center-of-mass oscillation in the $j$-th direction and is smaller than unity at low pressures relevant to this study. We take the ansatz that the solutions of eqs.~(\ref{eq:motion}) are written in the following form:

\begin{align}
\begin{split}\label{eq:sol}
q_j =& \dfrac{q_{j0}}{2}A_j e^{i\Omega_j t} + c.c. \\
q_{j0}=&\sqrt{-\kappa_j/\xi_j^{(j)}}
\end{split}
\end{align}
where $q_{j0}$ is a time-independent scale factor and $A_j$ is a time-dependent complex amplitude of the dynamics. 

The time derivative of $A_j$ is written as 

\begin{eqnarray}
\Dot{A_j}=\Omega_j \kappa_j \dfrac{dA_j}{d\tau}
\end{eqnarray}


The terms relevant to eqs.~(\ref{eq:motion}) are then given by
\begin{align}
\label{eq:qdot}
\Dot{q}_j=&\dfrac{q_{j0}\Omega_j}{2}\left( \kappa_j \dfrac{dA_j}{d\tau}+iA_j\right)e^{i\Omega_jt}+c.c. \\
\Ddot{q}_j=&\dfrac{q_{j0}\Omega_j^2}{2}\left( \kappa_j^2\dfrac{d^2A_j}{d\tau^2}+2i\kappa_j\dfrac{dA_j}{d\tau}-A_j\right)e^{i\Omega_jt} \notag\\
&+c.c. \\
q_j^3=&\dfrac{3q_{j0}^3}{8}|A_j|^2A_je^{i\Omega_jt}+c.c.\\
\epsilon_jq_j\cos\omega_jt =& \dfrac{\epsilon_jq_{j0}}{4}A_j^*e^{i\Omega_jt}e^{i\left(\omega_j-2\Omega_j\right)t}+c.c.\\
\label{eq:qsqq}
q_j^2q_k=&\dfrac{q_{j0}^2q_{k0}}{8}\left(A_j^2A_k^*e^{i\left(2\Omega_j-\Omega_k\right)t}+2|A_j|^2A_ke^{i\Omega_kt}\right) \notag \\ &  +c.c.
\end{align}
where $c.c.$ denotes complex conjugates and we ignored fast oscillating terms. The last term (\ref{eq:qsqq}) represents the cross-dimensional coupling that is introduced in the present study. Similarly to ref.~\cite{gieseler2013thermal}, we define rescaled parameters for $\Gamma_0$ and $\epsilon_j$:
\begin{align}
\begin{split}
\label{eq:rescale}
\Tilde{\gamma_0}&\equiv\dfrac{\Gamma_0}{\Omega_j\kappa_j} \\
\Tilde{\epsilon_j}&\equiv\dfrac{\epsilon_j}{\kappa_j}
\end{split}
\end{align}

Substituting eqs.~(\ref{eq:qdot}--\ref{eq:rescale}) into eq.~(\ref{eq:motion}) and considering the terms proportional to $\exp\left(i\Omega_j t\right)$ yields 
\begin{align}
\begin{split}
&\dfrac{q_{j0}\Omega_j^2}{2}\left(\kappa_j^2\dfrac{d^2A_j}{d\tau^2}+2i\kappa_j\dfrac{dA_j}{d\tau}-A_j\right) \\
&+\dfrac{q_{j0}\kappa_j\Omega_j^2}{2}\left(\kappa_j\dfrac{dA_j}{d\tau}+iA_j\right)\Tilde{\gamma_0} \\
&+\dfrac{q_{j0}\Omega_j^2}{4}\left(2A_j+\Tilde{\epsilon_j}\kappa_jA_j^*e^{i\left(\omega_j-2\Omega_j\right)t} \right) \\
&+\dfrac{q_{j0}\Omega_j^2}{4}\sum_{k\neq j}q_{k0}^2\xi_k^{(j)}|A_k|^2A_j\\
&+\dfrac{3}{8}q_{j0}^3\Omega_j^2\xi_j^{(j)}|A_j|^2A_j =0
\end{split}
\end{align}
where we assumed that $2\Omega_j$ is sufficiently separated from $\Omega_k$ for any combination of $\{j,k\}$. This is the case with the present study.

Considering that $\kappa_j$ is smaller than unity at high vacuum, we ignore higher order terms of $O(\kappa_j^2)$ and obtain

\begin{align}
\label{eq:motiona}
\begin{split}
\dfrac{dA_j}{d\tau}=&-\dfrac{\Tilde{\gamma_0}}{2}A_j+i\dfrac{\Tilde{\epsilon_j}}{4}A_j^*e^{i\left(\omega_j-2\Omega_j\right)t} \\
&+i\sum_{k\neq j}\dfrac{q_{k0}^2\xi_k^{(j)}}{4\kappa_j}|A_k|^2A_j-i\dfrac{3}{8}|A_j|^2A_j
\end{split}
\end{align}

Here we rewrite $A_j$ as 

\begin{align}
\label{eq:aredef}
A_j=p_j\exp\left(i\phi_j-i\delta_j\tau/2\right)
\end{align}
with $\delta_j=(2-\omega_j/\Omega_j)/\kappa_j$. Substituting eqs.~(\ref{eq:aredef}) into eqs.~(\ref{eq:motiona}), we obtain

\begin{align}
\begin{split}
\dfrac{dp_j}{d\tau}+ip_j\left(\dfrac{d\phi_j}{d\tau}-\dfrac{\delta_j}{2}\right)=&-\dfrac{\Tilde{\gamma_0}}{2}p_j+i\dfrac{\Tilde{\epsilon_j}p_j}{4}e^{-2i\phi_j} \\ &+i\sum_{k\neq j}\dfrac{q_{k0}^2\xi_k^{(j)}}{4\kappa_j}p_k^2p_j-i\dfrac{3}{8}p_j^3
\end{split}
\end{align}

The real and imaginary parts of these equations can be separately written as 

\begin{align}\label{eq:amp}
\dfrac{dp_j}{d\tau}=&-\dfrac{\Tilde{\gamma_0}}{2}p_j+\dfrac{\Tilde{\epsilon}_j}{4}p_j\sin2\phi_j \\
\begin{split}\label{eq:phase}
\dfrac{d\phi_j}{d\tau}=&\dfrac{\delta_j}{2}+\dfrac{\epsilon_j}{4}\cos2\phi_j+\sum_{k\neq j}\dfrac{q_{k0}^2\xi_k^{(j)}}{4\kappa_j}p_k^2-\dfrac{3}{8}p_j^2
\end{split}
\end{align}

For recovering equations without rescaling, we apply the inverse of the rescaling (\ref{eq:rescale}) to eqs.~(\ref{eq:amp},\ref{eq:phase}) and arrive at equations for the amplitude $Q_j=q_{j0}p_j$ and the phase $\phi_j$:

\begin{align}
\dfrac{dQ_j}{dt}=&-\dfrac{\Gamma_0}{2}Q_j+\dfrac{\epsilon_j\Omega_j}{4}Q_j\sin2\phi_j \\
\begin{split}\label{eq:finphase}
\dfrac{d\phi_j}{dt}=&\Omega_j-\dfrac{\omega_j}{2}+\dfrac{\epsilon_j\Omega_j}{4}\cos2\phi_j\\
&+\sum_{k\neq j}\dfrac{\xi_k^{(j)}\Omega_j}{4}Q_k^2+\dfrac{3}{8}\Omega_j\xi_j^{(j)}Q_j^2
\end{split}
\end{align}

The actual trap frequency $\omega_j/2$, which takes nonlinearities into account, is obtained by considering the steady state solution of eqs.~(\ref{eq:finphase}), namely, $d\phi_j/dt=0$ at weak parametric driving $\epsilon_j\approx 0$:

\begin{align}\label{eq:actfreq}
\dfrac{\omega_j}{2}=\Omega_j+\sum_{k\neq j}\dfrac{\xi_k^{(j)}\Omega_j Q_k^2}{4}+\dfrac{3}{8}\xi_j^{(j)}\Omega_j Q_j^2
\end{align}

\begin{figure}[t]
\includegraphics[width=0.95\columnwidth] {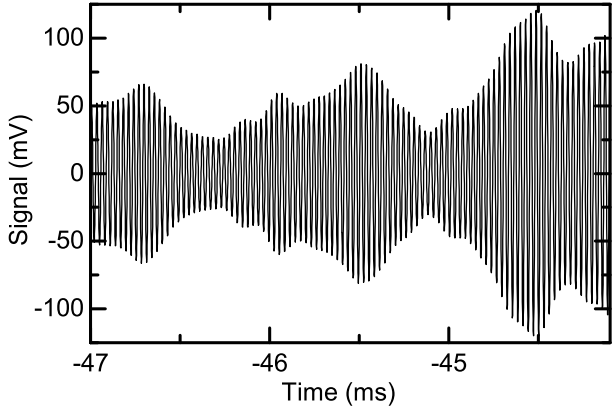}
\caption{Time trace of the center-of-mass oscillation of trapped nanoparticles in the $x$ direction at $\unit[0.6]{Pa}$. The amplitude of the oscillation strongly fluctuates. Due to the nonlinearities of the system, the amplitude fluctuation causes fluctuations in the trap frequencies not only in the $x$ direction but also in the $y$ and $z$ directions. The time scale of the amplitude fluctuation is faster at higher pressures.}
\label{fig:oscillation}
\end{figure}

These equations indicate that the actual trap frequency is lowered by finite oscillation amplitudes in any direction. The second terms in the right hand side of eqs.~(\ref{eq:actfreq}) represent the cross-dimensional couplings, whereas the third term represents the anharmonicity of the trap. Due to collisions with background gases, the oscillation amplitudes fluctuate (Fig.~\ref{fig:oscillation}). Amplitude fluctuations in a specific direction broadens the widths in the orthogonal two directions through the cross-dimensional couplings as well as the width in this direction through the anharmonicity. In reality, the oscillation amplitude fluctuates in all directions; therefore, the spectral widths are influenced by the motion in all directions. 

Our goal is to relate the temperature of the center-of-mass oscillation $T_{NP}$ to the spectral broadening via nonlinear frequency shifts in eqs.~(\ref{eq:actfreq}). The oscillation amplitude in the $j$-th direction is written by

\begin{align}\label{eq:thamp}
Q_j=\sqrt{\dfrac{2K_j}{m\Omega_j^2}}
\end{align}
where $K _j$ denotes the temporal center-of-mass energy of nanoparticles in the $j$-th direction.  The energies $K_j$ are expected to follow the Boltzmann distribution, which we experimentally confirm by taking the histogram of the observed signal (Fig.~\ref{fig:hist}). Substituting eq.~(\ref{eq:trapfreq}) and eqs.~(\ref{eq:thamp}) into eqs.~(\ref{eq:actfreq}), with nonlinear coefficients of Table~\ref{tab:nlc}, we obtain expressions for the temporal trap frequency in the $j$-th direction:

\begin{figure}[t]
\includegraphics[width=0.95\columnwidth] {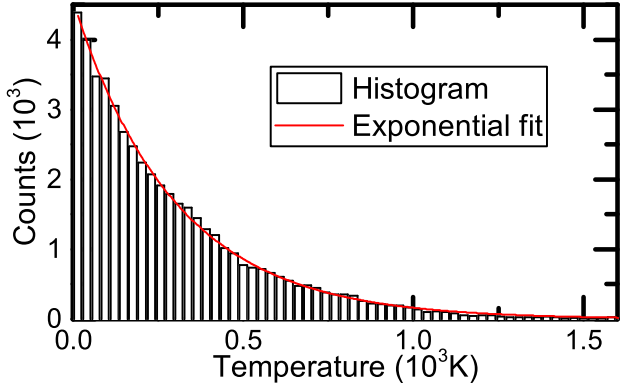}
\caption{(color online). Histogram of the energy of a single trapped nanoparticle in the $z$ direction. The horizontal axis is normalized with the assumption that the center-of-mass temperature of the nanoparticle is $\unit[300]{\rm K}$. We observe a clear exponential form consistent with the Boltzmann distribution.}
\label{fig:hist}
\end{figure}

\begin{align}\label{eq:osctemp}
\dfrac{\omega_j}{2}=&\Omega_j\left[1-\dfrac{1}{4U_0}\left(\sum_{k\neq j}K_k+\dfrac{3}{2}c_{j}K_j\right)\right]\notag\\
=&\Omega_j\left[1-\dfrac{E}{4U_0}\left(\dfrac{3c_j+4}{6}\right)\right]
\end{align}
where $E$ denotes the total center-of-mass energy and numerical factors $c_{j}$ is $2/3$ for $j=z$ and is $1$ for other cases. Here we assumed that the energy is equally distributed among each direction. The probability distribution of $E$ follows the Maxwell-Boltzmann distribution in three dimensions:

\begin{align}\label{eq:enfunc}
f_{\rm MB}(E) \propto\sqrt{E} \exp\left( -\dfrac{E}{k_BT_{NP}}\right)
\end{align}
From eqs.~(\ref{eq:osctemp},\ref{eq:enfunc}), we obtain the spectral profile of the thermal fluctuation for $\Delta \Omega_j = \omega_j/2-\Omega_j (<0)$

\begin{align}\label{eq:thdist}
f_{\rm th}(\Delta \Omega_j)\propto\sqrt{-\Delta \Omega_j}\exp\left(\dfrac{24U_0\Delta \Omega_j}{(3c_j+4)k_BT_{NP}\Omega_j}\right)
\end{align}
The peak of this function lies at 

\begin{align}\label{eq:peak}
\Delta \Omega_{j0} = -(3c_j+4)\dfrac{k_BT_{NP}}{48U_0}\Omega_j
\end{align}
For a small displacement near the peak $\delta = \Delta \Omega_j - \Delta \Omega_{j0}$, eq.~(\ref{eq:thdist}) is approximated by a Gaussian function:

\begin{align}\label{eq:thgauss}
f_{\rm th}(\delta) \propto \exp\left[ -\left( \dfrac{24U_0\delta}{(3c_j+4)k_BT_{NP}\Omega_j}\right)^2\right]
\end{align}

Thus, we arrive at a representation for the spectral width:

\begin{align}\label{eq:width}
\Delta \Omega_j \approx (3c_j+4)\dfrac{k_BT_{NP}}{24U_0}\Omega_j 
\end{align}
We anticipate that actual spectral profiles are given by the Lorentzian  function~\cite{gieseler2012subkelvin,vovrosh2017parametric}

\begin{align}\label{eq:psdlor}
S_{\rm Lor}(\omega)=\dfrac{2k_BT_{air}}{\pi m}\dfrac{\Gamma_0}{(\omega^2-\Omega_j^2)^2+\omega^2\Gamma_0^2}
\end{align}
weighted by the thermal distribution of eq.~(\ref{eq:thdist}).

\section{Comparison of experiments and theory}\label{sec:results}

\begin{figure*}[t]
\includegraphics[width=1.9\columnwidth] {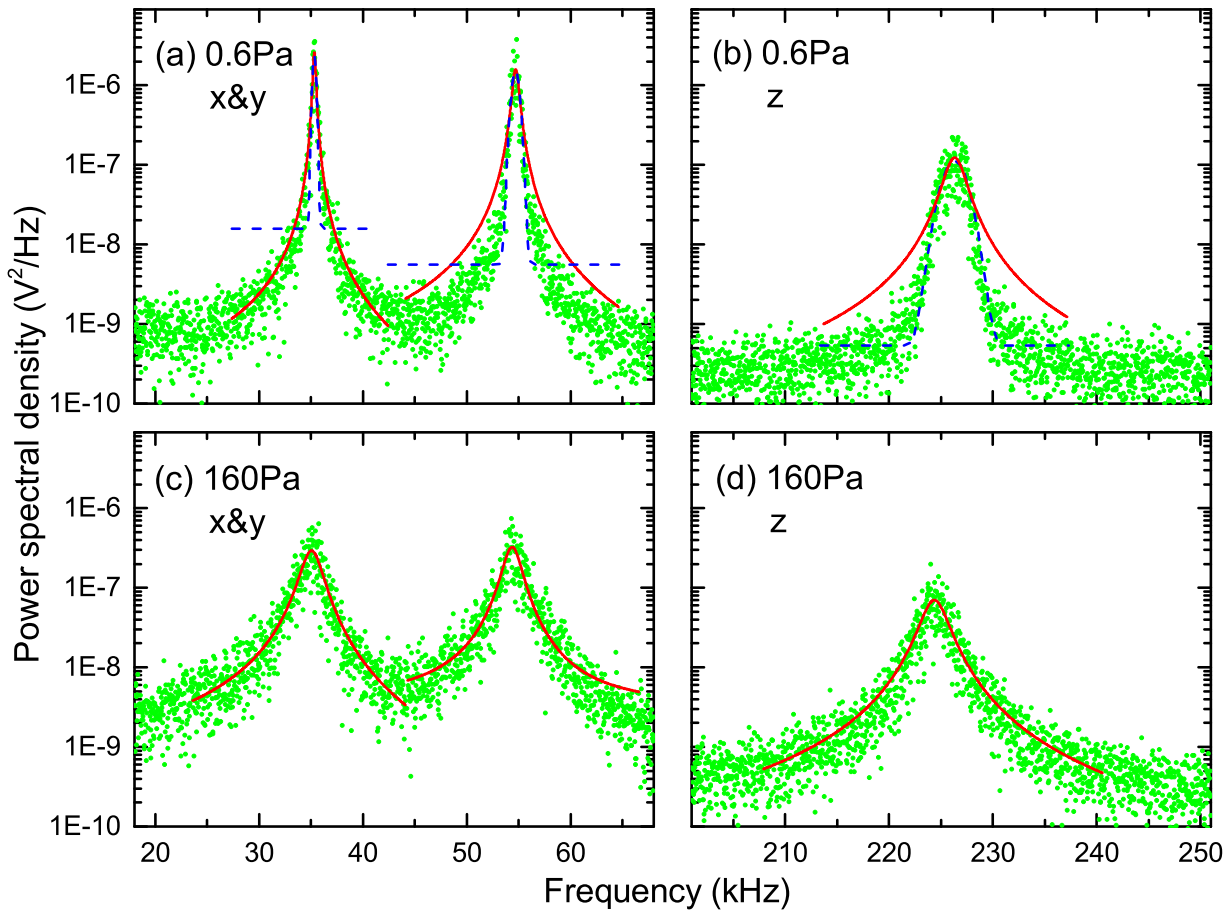}
\caption{(color online). PSD of a single nanoparticle with a fluorescence signal of $\unit[0.49]{\mu W}$. (a) For $x$ and $y$ directions at $\unit[0.6]{Pa}$, (b) for $z$ direction at $\unit[0.6]{Pa}$, (c) for $x$ and $y$ directions at $\unit[160]{Pa}$, and (d) for $z$ directions at $\unit[160]{Pa}$. The red solid lines are fits to the data with eq.~(\ref{eq:psdlor}), whereas the blue dashed lines are fits to the data with a Gaussian function. At high pressures, a fit with eq.~(\ref{eq:psdlor}) describes the observed spectra very well. At low pressures, a fit with eq.~(\ref{eq:psdlor}) significantly deviate and the Gaussian profiles better fit. The spectral width in the $z$ direction is much broader than those in the other directions at low pressure.}
\label{fig:spectra}
\end{figure*}

\begin{figure*}[t]
\includegraphics[width=1.9\columnwidth] {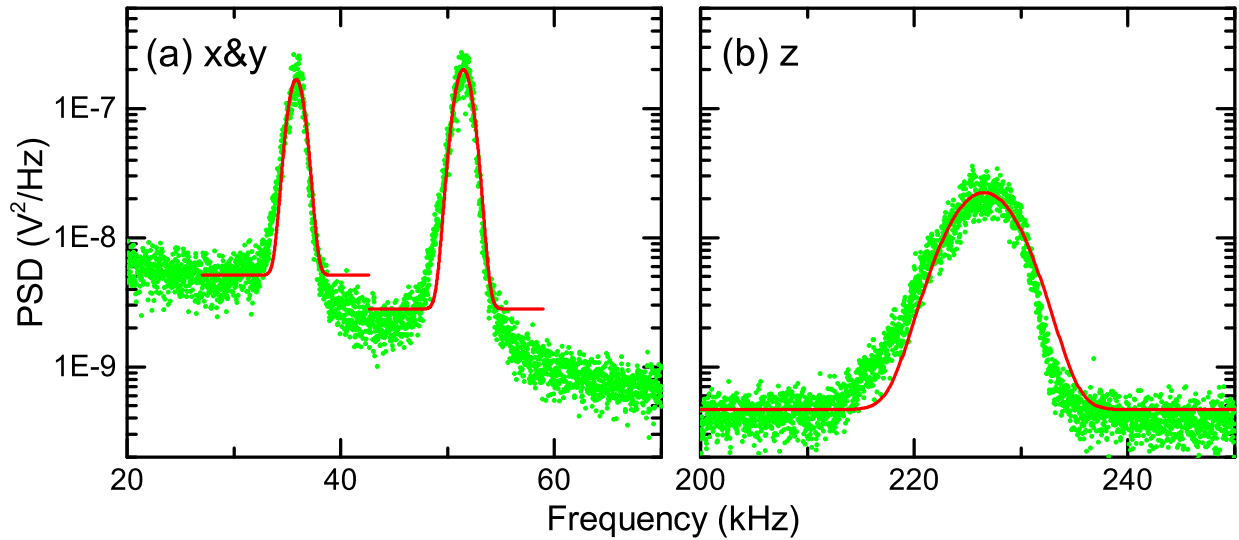}
\caption{(color online). PSD of a single nanoparticle with a fluorescence signal of $\unit[47]{nW}$ at $\unit[0.6]{Pa}$. (a) For $x$ and $y$ directions and (b) for $z$ direction. The spectral profiles are asymmetric and are closer to Gaussian functions than the Lorentzian functions for any directions. The asymmetric profiles indicate that the broadenings are dominantly of thermal origin. }
\label{fig:spectrasmall}
\end{figure*}

\begin{figure}[t]
\includegraphics[width=0.95\columnwidth] {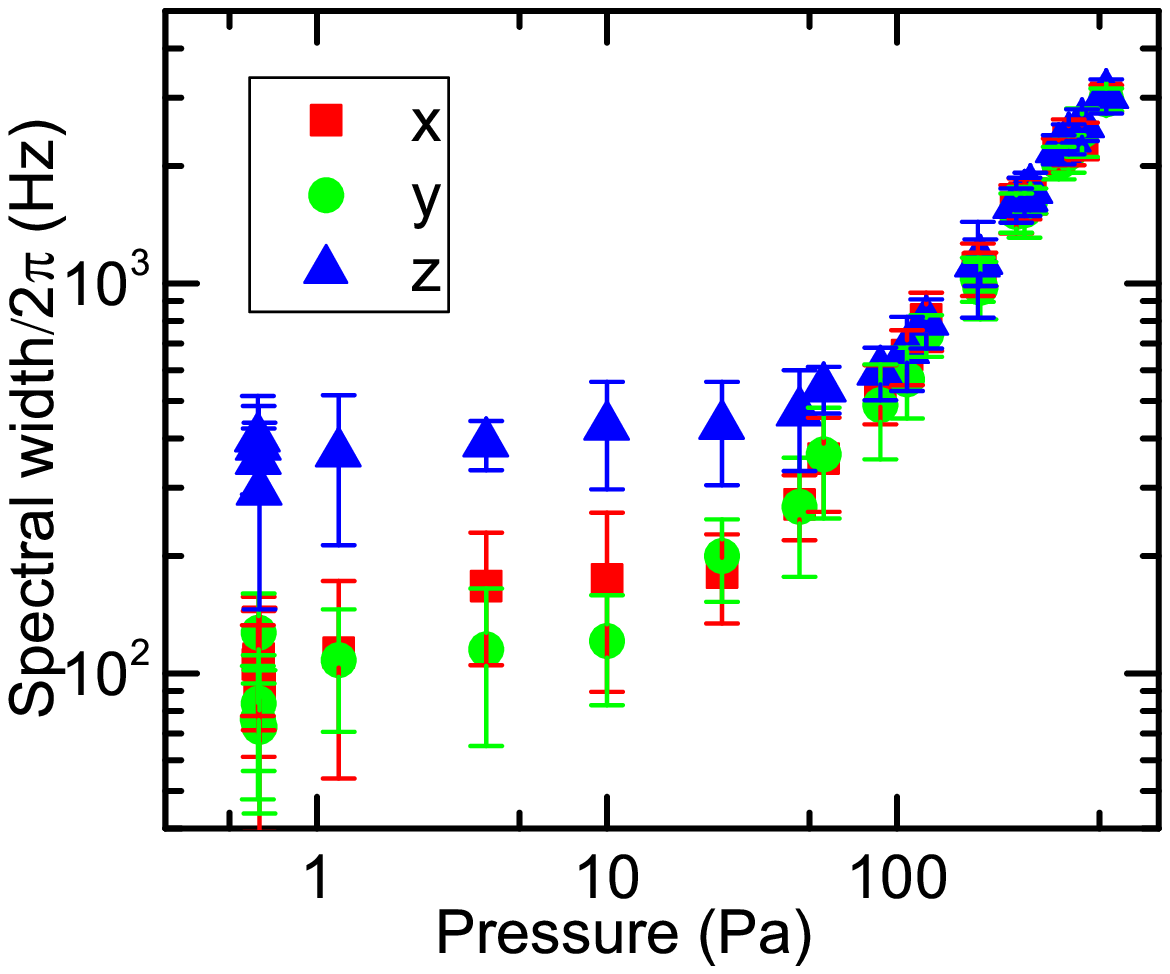}
\caption{(color online). Variation of the spectral widths with respect to the pressure for a single nanoparticles with a fluorescence signal of $\unit[25]{\mu W}$. The widths are obtained by fitting eq.~(\ref{eq:psdlor}) to the spectra. The widths linearly scale with the pressure and are independent on directions at high pressures, while at low pressures they settle to nearly constant values and are dependent on directions. The widths in $z$ direction are much broader than those in $x$ and $y$ directions. The error bars are the standard deviations for five data sets.}
\label{fig:pressure}
\end{figure}

We investigate the PSD of laser-trapped nanoparticles in a pressure range between 0.6 and $\unit[600]{Pa}$. The observed PSDs for a fluorescence signal of $\unit[0.49]{\mu W}$ are shown in Fig.~\ref{fig:spectra}. At high pressures above tens of Pa, the spectra fit very well with the Lorentzian function for any directions regardless of the size of nanoparticles. However, at lower pressures and for small nanoparticles, the observed profiles deviate from the Lorentzian function. The deviation is the most pronounced in the $z$ direction (Fig.~\ref{fig:spectra}), where the profile is nearly a Gaussian function. For even smaller nanoparticles, we observe asymmetric near-Gaussian profiles for three directions (Fig.~\ref{fig:spectrasmall}). Profiles for very large nanoparticles (with fluorescence signals of the order of $\unit[10]{\mu W}$) fit well with the Lorentzian function for any directions. 

The dependence of the observed width on the pressure is shown in Fig.~\ref{fig:pressure}. At pressures above tens of Pa, where the spectra are well described by the Lorentzian function, the widths are nearly identical among three directions and are proportional to the pressure. Such a behavior indicates the damping of nanoparticles' motion by collisions with background gases and is consistent with eq.~(\ref{eq:gamma}). At low pressures, where the profiles deviate from the Lorentzian function and are closer to Gaussian functions, the widths settle to values that are nearly independent from the pressure. Here the widths in the $z$ direction are much broader than those in the $x$ and $y$ directions.

Our observations are understood as follows. At high pressures, broadenings dominantly arise from the damping (\ref{eq:gamma}) and therefore the profiles agree well with the Lorentzian function. However, at low pressures the contribution of the damping (\ref{eq:gamma}) becomes smaller and that of the thermal fluctuation (\ref{eq:width}) is dominant. As a result, the profiles are closer to the Gaussian function (\ref{eq:thgauss}). Because the broadening due to the thermal fluctuations (\ref{eq:width}) is proportional to the trap frequency, the widths in the $z$ direction, in which the trap frequency is much higher, are much broader than those in $x$ and $y$ directions.

A previous work reported that taking a short duration for Fourier transform revealed an original narrow Lorentzian profile, while a long duration for Fourier transform resulted in a broad spectrum~\cite{gieseler2013thermal}. However, in our experiments, the spectral widths are nearly independent from the duration for Fourier transform. In the previous work, they focused on the thermal behavior in a specific direction and applied feedback cooling in the other two directions. By contrast, we investigate the situation without feedback cooling, implying that the cross-dimensional couplings broaden the spectral width even with a short duration for Fourier transform. 

\begin{figure}[t]
\includegraphics[width=0.95\columnwidth] {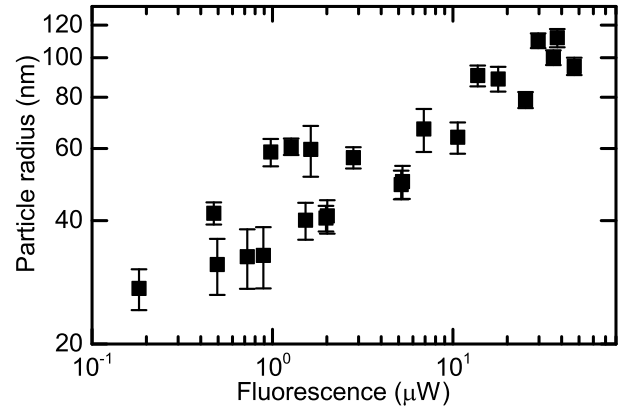}
\caption{Nanoparticles' radii with respect to the fluorescence signal. From the slope of the pressure dependence in Fig.~\ref{fig:pressure}, we derive radii of nanoparticles according to eq.~(\ref{eq:gamma}). The error bars indicate the standard deviation of the fit values in three directions. }
\label{fig:slope}
\end{figure}

In the following discussions, we focus on how the thermal broadening depends on the size of nanoparticles. We first find a correspondence between the nanoparticles' radii and the fluorescence signal. We convert the slope of the pressure dependence at high pressures into nanoparticles' radii by using eq.~(\ref{eq:gamma}). Fig.~\ref{fig:slope} plots the derived radii with respect to the fluorescence signal. The observed radii range between 20 and $\unit[120]{nm}$ and their mean value is in consistent with the specification of our sample ($\unit[60]{nm}$). The observed radii have large scatters, which we interpret as follows. The fluorescence signal is based only on radii, whereas the radii obtained from the slope depends also on the density of nanoparticles. Because we are interested in the interaction of nanoparticles with the trapping beam, hereafter we employ the fluorescence signal as a measure of the size of nanoparticles.

\begin{figure}[t]
\includegraphics[width=0.95\columnwidth] {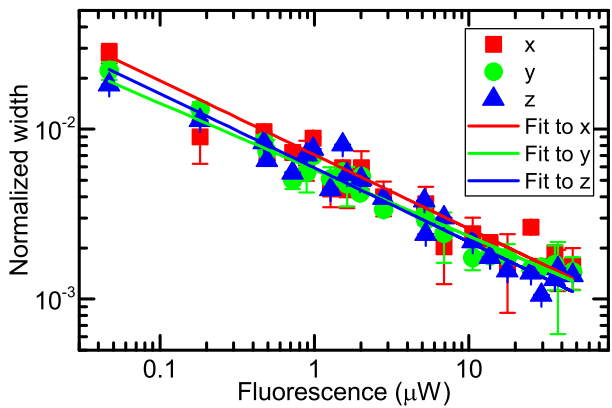}
\caption{(color online). Thermal widths normalized by the trap frequencies with respect to fluorescence values. The widths are obtained by fitting Gaussian profiles to the observed spectra. The solid lines indicate power-law fits to the data in each direction. The error bars indicate the standard deviation of the fit values from five data sets. Although the thermal widths are largely different among three directions, the widths normalized by the trap frequencies are nearly identical among three directions. }
\label{fig:normwidth}
\end{figure}

The spectral profiles at low pressures can be nearly Gaussian functions, for which fitting with the Lorentzian function may result in wrong fit values. For the systematic comparison of the observed widths among nanoparticles with largely different sizes, we fit the spectra at $\unit[0.6]{Pa}$ with Gaussian functions and extract the thermal widths for each. Although the extracted widths largely differ among three directions, we find that the widths normalized by the trap frequencies nearly fall on a single line in a wide range of the fluorescence signal (Fig.~\ref{fig:normwidth}). Such a behavior is in good agreement with our model in Sec.~\ref{sec:theory}. Although our model predicts a slight difference in widths of about $\unit[17]{\%}$ between $x,y$ and $z$ directions, we are unable to find a systematic difference among the three directions within the error of our measurements. 

Our model implies that the values of $k_BT_{NP}/U_0$ are directly obtained from the observed normalized widths. From Fig.~\ref{fig:normwidth}, we find that the ratio $k_BT_{NP}/U_0$ is ranged between $4\times 10^{-3}$ and $1\times 10^{-1}$. Assuming that the temperature of the center-of-mass motion $T_{NP}$ is $\unit[300]{K}$, we estimate that the trap depth $U_0$ is ranged between $k_B\times \unit[3\times 10^3]{K}$ and $k_B\times \unit[7\times 10^4]{K}$. For testing the validity of our model quantitatively, we compare the value of $U_0$ estimated from the spectral widths with that from a calculation based on the polarizability. On the one hand, at a mean radius of $\unit[60]{nm}$, we estimate that $U_0 \approx k_B\times \unit[1.5\times 10^4]{K}$ from the width. On the other hand, the trap depth is related to the intensity of the trap beam $I_0$ as~\cite{chang2010cavity}

\begin{align}\label{eq:potdepth}
U_0 = \dfrac{2\pi a^3}{c} {\rm Re}\left(\dfrac{\epsilon_r-1}{\epsilon_r+2}\right)I_0
\end{align}
where $c$ is speed of light and $\epsilon_r = 2.1$ is the dielectric constant of silica at $\unit[1550]{nm}$. With our beam intensity of $I_0=\unit[21]{MW/cm^2}$, the trap depth is calculated to be $k_B\times \unit[1.7\times 10^4]{K}$. Thus, our estimation from the spectral width is in good agreement with the purely calculated value.

The power-laws obtained from the fit in Fig.~\ref{fig:normwidth} are -0.44(1), -0.39(1), and -0.44(2) in $x$, $y$, and $z$ directions, respectively. According to our model, the thermal broadenings are proportional to $k_BT_{NP}/U_0$, which scales as $a^{-3}$ because of the scaling of $U_0 \propto a^3$ as shown in eq.~(\ref{eq:potdepth}). The fluorescence signal is expected to scale as $a^{6}$ because the scattering of the trapping beam by nanoparticles is in the regime of Rayleigh scattering ($a\ll \lambda$). Thus, theoretically we expect that the power-law of the normalized widths with respect to the fluorescence signal is $-0.5$. This value is in reasonable agreement with our observation. 

The slight deviation in the power-laws can, in principle, arise from laser absorption for the following reason. The larger the nanoparticles, the higher the internal temperature of laser-trapped nanoparticles, because the absorption scales with $a^3$ while cooling by surrounding gases scales with $a^2$. As reported in ref.~\cite{millen2014nanoscale}, when the internal temperature of a nanoparticle is higher than the temperature of the surrounding gases, it heats the colliding molecules and eventually heats the center-of-mass motion of nanoparticles as well via the back-action. Therefore, the larger the nanoparticles, the higher the center-of-mass temperature. This means $k_BT_{NP} \propto a^n$ with $n$ being positive and alters the power-laws in Fig.~\ref{fig:normwidth}.



\section{Conclusion}
We investigate the PSD of nanoparticles trapped in one-dimensional optical lattice in vacuum. At high pressures, the spectra are in good agreement with the profile expected from the model based on the damping with background gases. In this pressure regime, the widths linearly scales with the pressure. However, at pressures lower than tens of Pa, the spectra deviate from such a profile and become closer to a Gaussian profile. Here the widths do not scale with the pressure and nearly saturate to values specific to the orientation and the size of nanoparticles. For the $z$ direction, in which a standing wave is formed, the widths are much broader than those in the other two directions. We develop a theory to explain the widths at low pressures as broadenings originating from the thermal motions in three directions. We experimentally confirm the two important aspects of our model. First, the widths normalized by the trap frequency are nearly identical among three directions. Second, the normalized widths directly reveals the ratio of the thermal energy of nanoparticles to the trap depth. 

Our results are readily extended to the general case of a single-beam optical trap only with a slight modification on nonlinear coefficients. Furthermore, we believe that our results are generally the case with smaller particles such as atoms and molecules. The agreement between theory and experiments implies that the trap frequency is shifted by eq.~(\ref{eq:peak}) due to the thermal motion. Such a shift has been an important issue in the precision measurement with ultracold atomic gases~\cite{makhalov2015precision}.

In the present study, the PSD is fit either by the Lorentzian function or by the Gaussian function for clarity. Investigating the combined profile of the Lorentzian and Gaussian functions in the intermediate situation will be important future work. The detailed analysis of the spectral profile itself can provide the information on the nanoparticles' size. 

We expect that parametric feedback cooling of the center-of-mass motion~\cite{li2013millikelvin,gieseler2012subkelvin,millen2015cavity,fonseca2016nonlinear,jain2016direct,vovrosh2017parametric}, which can lower the temperature to a millikelvin regime, allows more elaborate studies on the relation between the widths and the thermal energy. Once the temperature is lowered to such a regime, the contribution of the thermal broadening to the widths is expected to be of the order of $\unit[1]{mHz}$ and is negligibly smaller than the broadening due to the damping. 

The method of studying the widths with respect to the size of nanoparticles is generally applicable to nanoparticles of other species. The widths at low pressures provide the important information on the ratio $k_BT_{NP}/U_0$, which is useful not only to estimate the trap depth but also to observe if $T_{NP}$ is dependent on the size of nanoparticles. Indirectly, the information on $T_{NP}$ tell us the extent of heating due to laser absorption.

\begin{acknowledgments}
We thank S.\,Inouye for lending us a fiber laser and M.\,Kozuma for fruitful discussions and also for lending us a diode-pumped solid state laser at the early stage of the experiments. We are grateful to Y.\,Matsuda for careful reading of the manuscript and R.\,Nagayama for experimental assistance. This work is supported by The Murata Science Foundation, The Mitsubishi Foundation, the Challenging Research Award and the "Planting Seeds for Research" program from Tokyo Institute of Technology, Research Foundation for Opto-Science and Technology, JSPS KAKENHI (Grant Number 16K13857 and 16H06016), JST PRESTO (Grant Number JPMJPR1661), and JST CREST (Grant Number JPMJCR16N4).
\end{acknowledgments}

\bibliographystyle{apsrev}


\end{document}